# Transport properties of graphene in the high-current limit

Amelia Barreiro[1], Michele Lazzeri[2], Joel Moser[1], Francesco Mauri[2], Adrian Bachtold[1]

1 CIN2(CSIC-ICN), Campus UAB, E-08193 Barcelona, Spain

2 IMPMC – Université Pierre et Marie Curie-Paris 6, CNRS, F-75015 Paris, France

We present a detailed study of the high-current transport properties of graphene devices patterned in a four-point configuration. The current tends to saturate as the voltage across graphene is increased but never reaches the complete saturation as in metallic nanotubes. Measurements are compared to a model based on the Boltzmann equation, which includes electron scattering processes due to charged and neutral impurities, and graphene optical-phonons. The saturation is incomplete because of the competition between disorder and optical-phonon scattering.

PACS: 72.80.Rj, 72.10.Di, 73.50.Fq

Building electronic devices based on graphene has been motivating intense research for the past few years [1,2]. Field-effect transistors have been demonstrated with mobility as high as 2.5 x $10^5$ cm$^2$ V$^{-1}$ s$^{-1}$ [3,4]. Moreover, large on-off ratios have been tailored by structuring graphene into nanoribbons [5-7]. Graphene also holds promise as interconnect, since it can sustain very high current density (about $10^9$ Acm$^{-2}$) [8]. Such current-carrying capacities are several orders of magnitude higher than that of present-day interconnects. The reason is that graphene is mechanically robust, chemically inert and it can sustain harsh conditions. Recently, it has been shown that the current saturates as the source-drain voltage is increased to a few volts [9]. The current saturation has been attributed to the scattering of electrons by surface phonons of the SiO$_2$ substrate (55 meV energy), and to a `pinch-off´ region where carriers flow through the Dirac point of graphene. These measurements have been carried out in devices structured in a two-point configuration and with a local gate.

In this Letter, we report on transport experiments in the high-current regime using a four-point configuration. The current $I$ tends to saturate as the voltage drop $V_{4pt}$ across the voltage electrodes increases, but the saturation is not complete (the differential conductance $dI/dV_{4pt}$ remains finite). In addition, $I$ in this regime can be tuned by sweeping the voltage $V_g$ applied to a gate. To understand these findings, we have computed the current using a Boltzmann approach. Good agreement is obtained using optical phonons of graphene (149 meV energy). In contrast to the case of carbon nanotubes, the high-field transport is very sensitive to the magnitude of elastic scattering, whereas hot phonon processes have a very negligible contribution.

Graphene flakes are obtained by mechanical exfoliation of Kish graphite on silicon wafers coated with a 280 nm thick silicon oxide layer. Single layers are identified using optical microscopy (for which the reflected light intensity has been calibrated by Raman spectroscopy (Fig. 1(a)). Graphene layers are patterned in a Hall-bar configuration using $O_2$ plasma etching. Cr/Au electrodes are fabricated using electron-beam lithography. Devices are annealed at 350 C in $H_2$/Ar for a few hours to remove fabrication residues. The current-induced cleaning technique is then employed in vacuum to increase the mobility [8]. Figure 1(b) shows an atomic force microscopy image of a device at the end of the fabrication process.

We employ a four-point configuration to avoid the contribution of the contact resistance at the graphene-electrode interfaces (Fig. 1(c)). This is important for the study of *I-V* characteristics as the contact resistance may change when increasing the applied voltage [10,11]. To preserve the integrity of the device in the high-current regime, measurements are carried out in vacuum ($10^{-6}$ mbar) and *I-V* characteristics are measured rapidly (within ~1 s). After each measurement, we control that the gate voltage dependence of the zero-field conductivity has not changed (Fig. 2(a)).

Figure 2(b) shows the *I-$V_{4pt}$* characteristics of a 350 nm wide graphene device. The current tends to saturate as $V_{4pt}$ increases. We however emphasize that the saturation is only a tendency (we have never observed for any device a full saturation where the differential conductance $dI/dV_{4pt}$ is 0). The current $I_{sat}$ in the saturation regime divided by the width can be as high as ~1 µA/nm. This value is rather similar to what is obtained in carbon nanotubes. Indeed, $I_{sat}$ for metal single-wall nanotubes is ~20 µA, which corresponds to ~3 µA/nm for a (hypothetical) unfolded nanotube [12]. As for individual shells of multi-wall nanotubes, $I_{sat}$ is ~30 µA (for a diameter of ~10 nm) corresponding to ~1 µA/nm [13].

Unlike metal nanotubes, however, $I_{sat}$ of graphene devices can be changed by applying a gate voltage (Fig. 2(b)). For further analysis, it is convenient to quantify $I_{sat}$ (even though the saturation is not complete). For this, we choose a procedure where we define a reference value for $dI/dV_{4pt}$ (red dashed line in (Fig. 2(c)). Then, we look at the corresponding $V_{4pt}$ using the $dI/dV_{4pt}(V_{4pt})$ plot, which allows us to get $I_{sat}$ from the $I(V_{4pt})$ curve. This procedure is rather insensitive to the reference value of $dI/dV_{4pt}$ (the error bars in Fig. 2(d) quantify the effect of choosing different reference values).

Similar $I$-$V_{4pt}$ characteristics have been measured for 7 devices with different graphene widths $W$ and lengths $L$ (between the voltage electrodes). Figure 3(a) shows that $I_{sat}$ scales linearly with the width. However, $I_{sat}$ is quite insensitive to the length (Fig. 3(b)) which suggests that the voltage drops linearly along the channel.

We now discuss possible origins of the transport properties in the saturation regime. The current saturation could be related to the backscattering of electrons when they flow through the zero-energy band gap (pinch-off region). This is a well-documented phenomenon in MOSFETs [14] where $I_{sat}$ goes down to 0 for $V_g$ approaching the energy gap. However, this is clearly not observed in the measurements of Fig. 2b.

We consider the scattering of electrons by the optical phonons of graphene. This scattering is likely to be relevant as it is responsible for the current saturation in nanotubes [12]. When applying an electric field $\xi$, electrons are accelerated so they can gain the energy to emit an optical phonon of energy $\hbar\Omega$. When emitting a phonon, electrons are backscattered. The corresponding "activation" scattering length is $l_\Omega = L\hbar\Omega/(eV_{4pt})$ [12]. If the phonon emission is instantaneous and the elastic scattering (due to defects) is negligible, a steady-state population is established in which the right moving electrons are populated to an energy $\hbar\Omega$ higher than the left moving ones (Fig. 4(a)). The current at "full saturation" can be calculated by integrating the velocity along the channel axis over this steady state population. It reads:

$$I_{FS} = W \frac{2e}{\pi^2 \hbar^2 v_F} \left( \frac{C_g^2 \pi^2 V_g^2 (\hbar v_F)^4}{e^2 (\hbar\Omega)^2} + \frac{(\hbar\Omega)^2}{4} \right) \qquad \text{when } V_g < \frac{e(\hbar\Omega)^2}{2\pi C_g \hbar^2 v_F^2}$$

$$I_{FS} = W \frac{2e\hbar\Omega}{\pi^2 \hbar^2 v_F} \sqrt{\frac{C_g \pi V_g (\hbar v_F)^2}{e} - \frac{(\hbar\Omega)^2}{4}} \qquad \text{otherwise} \qquad (1)$$

where $v_F$ is the Fermi velocity and $C_g$ the gate-graphene capacitance is $C_g = 1.15 \cdot 10^{-4}$ F/m$^2$. These expressions have been obtained by imposing charge conservation, namely that the integration of charge over all relevant states remains equal to $C_g V_g/e$. The A'$_1$ phonon at **K** is the one with the largest electron-phonon coupling and has energy $\hbar\Omega$ =149 meV [15]. Fig. 2(d) reports the corresponding $I_{FS}$. The calculated full saturation-current overestimates the experiment, although being of the same order of magnitude.

The above model can be improved by including the effect of elastic scatterers. We assume conic bands with Fermi velocity $v_F=10^6$ m/s and we solved numerically the Boltzmann transport equation to obtain, in the presence of a uniform electric field, the steady-state population of the conduction electrons $f(\mathbf{k})$ and of the optical phonons $n^{\Gamma\text{-TO}}(\mathbf{k})$, $n^{\Gamma\text{-LO}}(\mathbf{k})$, $n^{\mathbf{K}}(\mathbf{k})$, where **k** is the momentum. $\Gamma$ and **K** refer to the two E$_{2g}$ and A'$_1$ branches near **Γ** and **K**. Populations are obtained from a coupled electron-phonon dynamic as in [16] without any further approximations than those inherent to the Boltzmann treatment. Here we suppose that populations are spatially homogeneous (*i.e.* they do not depend on the real-space coordinates), as supported from Fig. 3(b). All the intrinsic parameters are obtained from ab-initio calculations, validated against experimental temperature–dependent phonon dispersions and line-widths [15,17]. In particular, the magnitude of the electron-phonon couplings is obtained from the GW calculations of [15] with an angular-dependence as in Eqs. (6,7) of [18]. Optical phonons decay into acoustic phonons with anharmonic phonon-decay rates from [17] with the thermalisation mechanism described in [16], i.e. the temperature of acoustic phonons is supposed to remain much smaller that the Debye temperature of optical phonons (1740 K). The elastic electron-scattering is due to defects and it is an extrinsic property which depends on the specific sample. We suppose that electrons are scattered mainly by charged defects and neutral point-defects, with densities $n_{ic}$ and $n_{i\delta}$, respectively [19]. The form of the scattering rate is taken from [20]. We obtain $n_{ic}=0.41\times10^{12}$cm$^{-2}$ and $n_{i\delta}=0.27\times10^{12}$cm$^{-2}$ (model "Cδ", see Fig. 2a) by fitting the Boltzmann results to the measured zero-voltage conductivity σ [21], which is entirely determined by the transport elastic-scattering length $l_{el}$ at the Fermi level $\varepsilon_F$: $\sigma = \varepsilon_F l_{el} e^2 /(\pi \hbar^2 v_F)$. Also, we consider a second model (model "C") in which $n_{i\delta}=0$ and $n_{ic}$ depends on the gate voltage. Perfect agreement with the measured σ is obtained with $n_{ic}=(0.32+0.019|V_g|/\text{Volt})\times10^{12}$cm$^{-2}$ [21]. Since both models reproduce the measured σ equally well, they have the same $l_{el}$ at $\varepsilon_F$, but different scattering rates away from $\varepsilon_F$ with different scattering-angle dependences [20]. Thus, comparing the two models allows us to study the dependence of the high-field current on the specific form of the elastic scattering. We also considered the effect of electron-electron Coulomb scattering, which turns out to give minor corrections and is thus neglected in the following [22]. Finally, we restrict ourselves to cases in which the conduction is determined by only one type of carriers, namely electrons or holes. In fact, the present semi-classical treatment does not account for the production of electron-hole pairs, which is relevant for small $V_g$ values (below ~12 V) and it is beyond the scope of the present paper.

The calculated *I-V* curves for the two models "Cδ" and "C" at $V_g=\pm 24$ V are compared with experiments in Fig. 4(c). The computed saturation intensities as a function of $V_g$ are reported in Fig. 2(d). The calculated currents are somewhat lower than the measured one, but the agreement is rather satisfactory, especially considering that no fitting parameter has been employed beside the zero-field conductivity. The present model for carrier-defect scattering [20] is symmetrical for electrons and holes, thus the calculations for $V_g=\pm 24$ V are identical. On the contrary, the measured *I-V* curves (Fig. 4(c)) and saturation intensities (Fig. 2(d)) are not symmetrical by changing the sign of $V_g$. These asymmetries become more pronounced upon increasing $V_{4pt}$ at high $|V_g|$. This electron/hole asymmetry could be explained by the presence of a resonant defect scattering state well below the energy of the Dirac point [23].

The models "Cδ" and "C" account for the experimental finding that the saturation of *I* is never complete, which cannot be captured by the "full saturation" model, Eq.(1). The dependence of the results on the elastic-scattering model indicates that high-field measurements are a sensitive probe to the details of the elastic-scattering processes, contrary to the low-field ones which just depend on the average transport scattering length. In carbon nanotubes the high-field current is strongly affected by a large non-equilibrium population of optical phonons (hot-phonons) [16,18,24,25]. In the present case, the current is not affected by an increase in the optical-phonon population. Indeed, the phonon population $n$ remains very small (in all our simulations $n^{\Gamma\text{-TO}}(\mathbf{k})$, $n^{\Gamma\text{-LO}}(\mathbf{k})$, $n^{K}(\mathbf{k}) < 0.003$), the optical phonon scattering rate being proportional to $(1+2n)$ [18].

To understand the behavior of the *I-V* curve, in Fig. 4(d) we compare optical-phonon "activation" scattering-length $l_\Omega$ with the transport elastic scattering lengths $l_{el}$ and with the lengths traveled by electrons (once the phonon emission threshold has been reached) before phonon scattering, $l_{ph}^{\Gamma}$, $l_{ph}^{K}$, which depend on the electron-phonon coupling strength. Since $l_\Omega \propto 1/V_{4pt}$, at low field $l_{el} << l_\Omega$ and the elastic scattering is the dominating process. At high field ($V_{4pt} \sim 1$ V) $l_{el} \sim l_\Omega$, the optical-phonon emission is activated and the current starts to saturate. However, the full saturation, Eq. (1), will occur only if the phonon emission is instantaneous ($l_{ph}^{K} << l_\Omega$) and elastic scattering processes are negligible ($l_\Omega + l_{ph}^{K} << l_{el}$). This condition is never realized. Thus the electronic population does not vary abruptly, Fig.4(b), the current is below the full saturation limit (Eq. (1)) and continues to slightly increase upon increasing $V_{4pt}$.

Measurements in a previous work [9] have been interpreted using surface phonons of the SiO$_2$ substrate (55 meV energy). The introduction of this further phonon-scattering channel (in addition to that associated to the graphene optical-phonons, which are intrinsic to the system) would further decrease the current, providing a worse agreement with measurements.

In conclusion, we have investigated the current saturation in graphene devices. Our results show that the high field transport is very sensitive to the magnitude of elastic scattering and the current does not completely saturate. This is because, especially at the Dirac point (see Fig. 4(d)), the elastic scattering length (~60 nm) is shorter than the electron-phonon scattering lengths (~1 μm). In contrast, metallic nanotubes have much longer elastic lengths (up to ~1 μm) and shorter electron-phonon lengths (~200 nm) [18], which are both independent from the energy. For this reason, the current through nanotubes has a kink at ~0.2 V and fully saturates at high field [12]. In graphene, a full saturation regime could be reached only by a reduction of the disorder (elastic scattering) with an increase of the mobility by more than one order of magnitude.

We thank N. Camara and H. Tao for experimental help. The research was supported by an EURYI Grant and FP6-IST-021285-2. Calculations were done at IDRIS (Orsay).

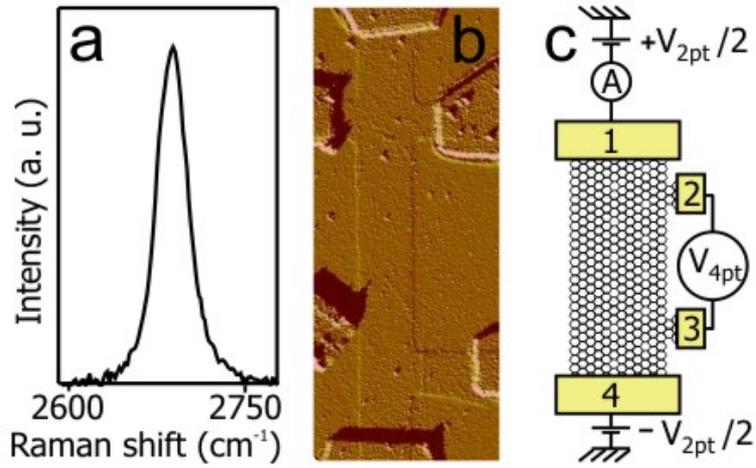

Fig. 1(a) Raman spectrum of a single layer. (b) Atomic force microscopy image of a graphene device. The channel width is 1μm. (c) Schematic of the four-point measurement. The device is symmetrically voltage biased. $I$ is measured through electrode 1 and $V_{4pt}$ between electrodes 2 and 3.

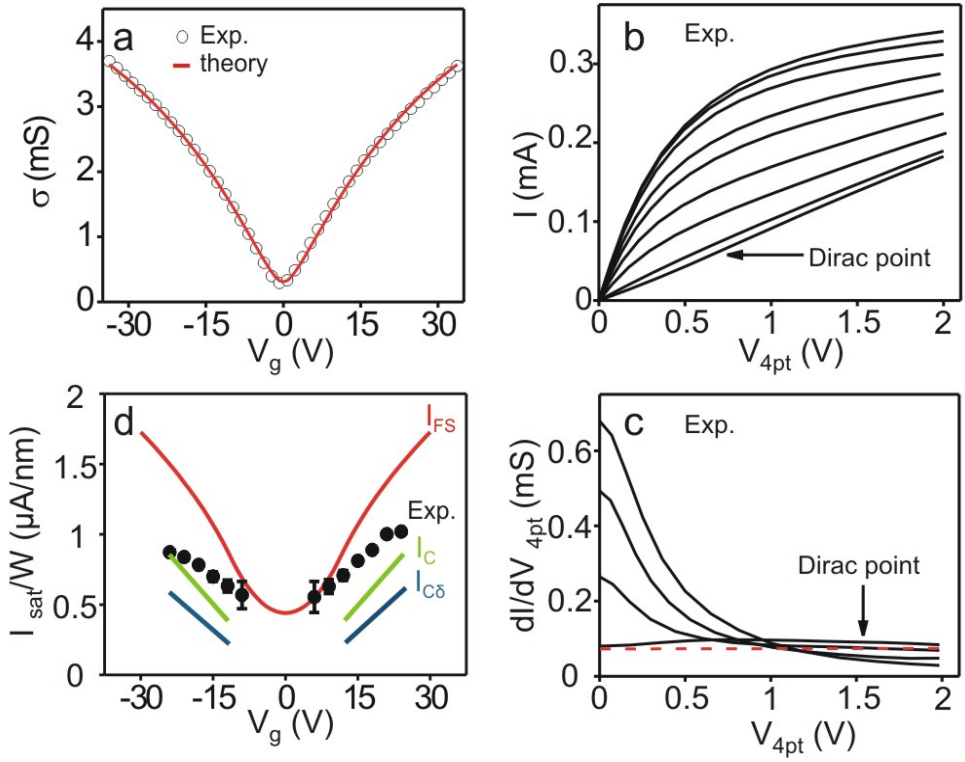

Fig. 2(a) Conductivity as a function of $V_g$ measured at zero voltage ($V_{4pt} < kT$) at 300 K and $10^{-6}$ mbar. $W = 350$ nm and $L = 1.3$ μm. The red curve is the theoretical calculation with the model "Cδ". Similar agreement is obtained with the model "C" (b) $I$-$V_{4pt}$ characteristics for different $V_g$ (from 0 V to -24 V in 3 V steps). (c) Numerically differentiated curves $dI/dV_{4pt}$ as a function of $V_{4pt}$ for different $V_g$ (from 0 V to -18 V in 6 V steps). The red dashed line is the reference value for the evaluation of $I_{sat}$. (d) $I_{sat}$ as a function of $V_g$. The red curve corresponds to the full saturation, Eq. (1), with $\hbar\Omega = 149$ meV. The green and blue curves are the results of the Boltzman calculation with the models "Cδ" and "C", respectively.

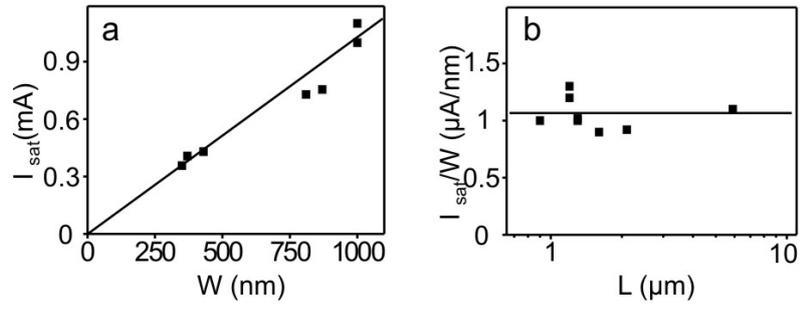

Fig. 3 (a,b) Current $I_{sat}$ as a function of width and length for 7 different graphene samples at $V_g$ about -25 V from the Dirac point. The mobility varies from 4500 to 15000 cm$^2$ V$^{-1}$ s$^{-1}$. The straight lines are guides to the eye.

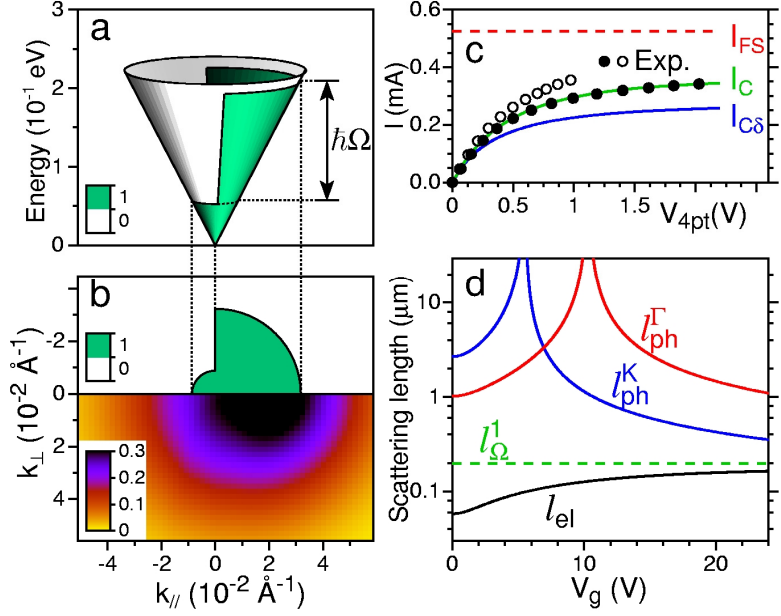

Fig. 4 (a) Electron population (green $f=1$, white $f=0$) on the Dirac cone in the ideal saturation regime (Eq. 1). (b) Electron population in the ideal saturation regime (upper half) and according to the Boltzmann treatment with the model "C$\delta$" (lower half) at $V_g=+24$ V and $V_{4pt}=2$ V. $\mathbf{k}_{//}$ is the direction parallel to the electric field and $\mathbf{k}=\mathbf{0}$ is the Dirac point. (c) Current as a function of voltage at $V_g=\pm24$ V. Lines are calculations done in the ideal saturation regime ($I_{FS}$) and with the two Boltzmann models ($I_C$, $I_{C\delta}$). Full (open) dots are measurements for $V_g=-24$ (+24) V. (d) Electron transport scattering lengths: elastic ($l_{el}$), electron-phonon ($l_{ph}$), and optical-phonon activation ($l_\Omega^1=198$ nm) at $V_{4pt}=1$ V. Notice that $l_\Omega$ varies with $V_{4pt}$ as $l_\Omega=l_\Omega^1$(Volt/$V_{4pt}$). $l_{el}$, $l_{ph}^\Gamma$, and $l_{ph}^K$ are obtained taking into account the angular dependence of the scattering rate as in Eq. (1) of [20]. $l_{ph}^\Gamma$ and $l_{ph}^K$ are computed for the electrons at $\varepsilon_F+\hbar\Omega/2$, supposing that the states at $\varepsilon_F-\hbar\Omega/2$ are empty (with $\hbar\Omega$ corresponding to the $\Gamma$ or $K$ phonon). Notice that $l_{ph}$ diverges for $\varepsilon_F - \hbar\Omega/2 = 0$ (in contrast to the case of nanotubes) since the density of final states vanishes at the Dirac point.